\documentclass{article}%
\usepackage{graphicx}
\usepackage{amsmath}
\usepackage{amsfonts}
\usepackage{amssymb}%
\newtheorem{theorem}{Theorem}
\newtheorem{acknowledgement}[theorem]{Acknowledgement}

\begin{document}

\title{{\large Two interacting spins in external fields.}\\{\large Four-level systems.}}
\author{V.G. Bagrov\thanks{On leave of absence from Tomsk State University and Tomsk
Institute of High Current Electronics, Russia, e-mail: bagrov@phys.tsu.ru},
M.C. Baldiotti\thanks{E-mail: baldiott@fma.if.usp.br}, D.M.
Gitman\thanks{E-mail: gitman@dfn.if.usp.br}, and A.D. Levin\thanks{Dexter
Research Center, USA; e-mail: SLevin@dexterresearch.com},\\Instituto de F\'{\i}sica, Universidade de S\~{a}o Paulo,\\Caixa Postal 66318-CEP, 05315-970 S\~{a}o Paulo, S.P., Brazil}
\maketitle

\begin{abstract}
In the present article, we consider the so-called two-spin equation that
describes four-level quantum systems. Recently, these systems attract
attention due to their relation to the problem of quantum computation. We
study general properties of the two-spin equation and show that the problem
for certain external backgrounds can be identified with the problem of one
spin in an appropriate background. This allows one to generate a number of
exact solutions for two-spin equations on the basis of already known exact
solutions of the one-spin equation. Besides, we present some exact solutions
for the two-spin equation with an external background different for each spin
but having the same direction. We study the eigenvalue problem for a
time-independent spin interaction and a time-independent external background.
A possible analogue of the Rabi problem for the two-spin equation is defined.
We present its exact solution and demonstrate the existence of magnetic
resonances in two specific frequencies, one of them coinciding with the Rabi
frequency, and the other depending on the rotating field magnitude. The
resonance that corresponds to the second frequency is suppressed with respect
to the first one.

\end{abstract}

\section{Introduction}

\subsection{Overview}

Finite-level systems have always played an important role in quantum physics.
In particular, two-level systems possess a wide range of applications, for
example, in the semi-classical theory of laser beams \cite{Nus73}, optical
resonance \cite{AllEb75}, absorption resonance, and nuclear induction
experiments \cite{RabRaS54}, in the explanation of the behavior of a molecule
in a cavity immersed in electric or magnetic fields \cite{FeyVe57}, and so on.
The best known physical systems that could be identified with two-level
systems are, for example, two-level atoms (atoms in which the interaction with
specific electromagnetic fields naturally selects only two energy levels
important for the consideration) and spin-one-half objects interacting with a
magnetic field. Four-level systems describe two interacting one-half spins,
e.g., those of an electron and a proton in an atom, or those of two electrons
frozen in space, and so on,\ see, e.g., \cite{Fey65,Tow92}. Recently, two- and
four-level systems attract even more attention due to their relation to the
problem of quantum computation, see, for example, \cite{QuantComp}. In this
problem, the state of each bit of conventional computation is permitted to be
any quantum-mechanical state of a \textit{qubit} (quantum bit), which can be
regarded as a two-level system. Computation is performed by manipulating these
qubits with the help of the so-called \textit{quantum gates}. Although these
gates depend on the number of involved qubits, it is possible to demonstrate
that all manipulations can be efficiently accomplished by using gates with
just one and two qubits, where two-qubit gates can be identified with a
four-level system, see \cite{QuantComp}. For these reasons, two- and
four-level systems are crucial elements of possible quantum computers. For
physical applications, it is very important to have explicit exact solutions
of two- and four-level system equations. One can mention, in this respect, the
Rabi solution of a two-level system equations (\textit{the spin equation}
according to our terminology), which has a great importance in the treatment
of numerous physical phenomena. In our opinion, the most recent and complete
study of the spin equation and its exact solutions is presented in our work
\cite{BagGiBL05}. In the present article, we turn our attention to four-level
systems. We study the general properties of the corresponding Schr\"{o}dinger
equation (called \textit{the two-spin equation} in what follows) and present
some exact solutions of this equation. In sec. 2, we define a two-spin
equation with an external field and describe its general properties. In
particular, we show that the problem for an external background equal for both
spins is reduced to the problem of one spin in a certain background. This
allows one to generate a number of exact solutions for the two-spin equation
on the basis of already known exact solutions of the spin equation. In sec. 3,
we present some exact solutions for the two-spin equation with external
backgrounds different for each spin but having the same direction. In sec. 4,
we study the eigenvalue problem for a time-independent spin interaction and a
time-independent external background. In sec. 5, we define a possible analogue
of the Rabi problem for the two-spin equation. We present its exact solution
and show the existence of magnetic resonances in two specific frequencies, one
of them coinciding with the Rabi frequency, and the other depending on the
rotating field magnitude. The resonance that corresponds to the second
frequency is suppressed with respect to the first one.

\subsection{Spin equation}

The nonrelativistic spin operator $\mathbf{\hat{s}}=\frac{\hbar}%
{2}\boldsymbol{\sigma}$ is a particular case of the momentum
operators\footnote{Here, $\mathbf{\sigma}=\left(  \sigma_{1},\sigma_{2}%
,\sigma_{3}\right)  $ are the Pauli matrices. In what follows, we set
$\hbar=1$.} and describes particles with spin $s$ one-half ($\mathbf{\hat{s}%
}^{2}=3/4=s\left(  s+1\right)  ,\;s=1/2$). Consider the $\vartheta_{\lambda}%
$-basis%
\begin{equation}
s_{z}\vartheta_{\lambda}=\left(  -1\right)  ^{\lambda-1}s\vartheta_{\lambda
}\,,\;\lambda=1,2\,;\;\vartheta_{1}=\left(
\begin{array}
[c]{c}%
1\\
0
\end{array}
\right)  \,,\;\vartheta_{2}=\left(
\begin{array}
[c]{c}%
0\\
1
\end{array}
\right)  \,.\label{a.1}%
\end{equation}
In the $\vartheta_{\lambda}$-basis, state vectors $\psi\left(  t\right)  $ can
be associated with a time-dependent two-columns:%
\begin{equation}
\psi\left(  t\right)  =\sum_{\lambda=1,2}v_{\lambda}\left(  t\right)
\vartheta_{\lambda}\Longleftrightarrow\psi\left(  t\right)  =\left(
\begin{array}
[c]{c}%
v_{1}\left(  t\right)  \\
v_{2}\left(  t\right)
\end{array}
\right)  .\label{a.2}%
\end{equation}
The dynamics of a spin-one-half particle subject to a time-dependent external
field is described by the Schr\"{o}dinger equation with a Hamiltonian $\hat
{h}$. In the $\vartheta_{\lambda}$-basis, the most general form of the
Hamiltonian is $\hat{h}=2\left(  \mathbf{\hat{s}\cdot F}\right)  =\left(
\boldsymbol{\sigma}\mathbf{\cdot F}\right)  $, where $\mathbf{F=}\left(
F_{1}\left(  t\right)  ,F_{2}\left(  t\right)  ,F_{3}\left(  t\right)
\right)  $ is an arbitrary time-dependent vector (external field). Then the
corresponding Schr\"{o}dinger equation reads%
\begin{equation}
i\frac{d\psi}{dt}=\hat{h}\psi\,,\;\psi=\left(
\begin{array}
[c]{c}%
v_{1}\left(  t\right)  \\
v_{2}\left(  t\right)
\end{array}
\right)  ,\;\hat{h}=\left(
\begin{array}
[c]{cc}%
F_{3} & F_{1}-iF_{2}\\
F_{1}+iF_{2} & -F_{3}%
\end{array}
\right)  \,.\label{a.3}%
\end{equation}
Equation (\ref{a.3}) implies the following coupled equations for the
components $v_{i}\left(  t\right)  $, $i=1,2$:%
\begin{equation}
i\dot{v}_{1}=F_{3}v_{1}+\left(  F_{1}-iF_{2}\right)  v_{2}\,,\;i\dot{v}%
_{2}=-F_{3}v_{2}+\left(  F_{1}+iF_{2}\right)  v_{1}\,.\label{1.8}%
\end{equation}
Equation (\ref{a.3}) is called \textit{the spin equation}. The spin
equation\ with a real external field can be regarded as a reduction of the
Pauli equation \cite{Pau27} to the ($0+1$)-dimensional case. Such an equation
is used to describe a (frozen in space)\ spin-$1/2$ particle of magnetic
momentum $\mu$, immersed in a magnetic field $\mathbf{B}$ (in this case,
$\mathbf{F}=-\mu\mathbf{B}$), and\textbf{ }has been intensely studied in
connection with the problem of magnetic resonances \cite{RabRaS54,BloSi40}.
The spin equation with complex external fields describes a possible damping of
two-level systems \cite{BagGiBL05}. There exist various equations that are
equivalent, or (in a sense) related, to the spin equation. For example, the
well-known top equation, which appears in the gyroscope theory, in the theory
of precession of a classical gyromagnet in a magnetic field (see
\cite{FeyVe57}), and so on. The spin equation with an external field in which
$F_{s}\left(  t\right)  ,\;s=1,2$ are purely imaginary and $F_{3}$ is constant
is a degenerate case of the Zakharov--Shabad equation, which plays an
important role in the soliton theory \cite{ZakSh84}. The first exact solution
of the spin equation was found by Rabi \cite{Rab37} for an external field of
the form $\mathbf{F}=\left(  f_{1}\cos\omega t,\,f_{2}\sin\omega
t,\,F_{3}\right)  $, where $f_{1,2},\,\omega,$ and $F_{3}$ are real constants.
A number of exact solutions of the spin equation were found in
\cite{BagGiBL05,BagBaGW01,BagBaGS04}. For periodic, or quasiperiodic, external
fields, the equations of a two-level system have been studied by many authors
using different approximation methods, e.g., perturbative expansions
\cite{BarCo02}, see also \cite{Bar00}.

\section{Two-spin equation and its properties}

\subsection{Two-spin equation}

Let us consider two interacting spins one-half. We choose the state space for
such a system as the direct-product space of the state spaces of individual
spins. In this space, we choose basis states $\Theta_{\mu}\,,$ $\mu=1,2,3,4,$
as the direct product of individual bases:%
\begin{equation}
\Theta_{1}=\vartheta_{1}\otimes\vartheta_{1}\,,\;\Theta_{2}=\vartheta
_{1}\otimes\vartheta_{2}\,,\;\Theta_{3}=\vartheta_{2}\otimes\vartheta
_{1}\,,\;\Theta_{4}=\vartheta_{2}\otimes\vartheta_{2}\,. \label{basis}%
\end{equation}
The spin operators for the first and second subsystems are $\mathbf{\hat{s}%
}_{1}$ and $\mathbf{\hat{s}}_{2},$%
\begin{align*}
&  \mathbf{\hat{s}}_{1}=\frac{\boldsymbol{\sigma}}{2}\otimes
I\,,\;\mathbf{\hat{s}}_{2}=I\otimes\frac{\boldsymbol{\sigma}}{2}\,,\\
&  \hat{s}_{1z}\Theta_{1,2}=\frac{1}{2}\Theta_{1,2}\,,\;\hat{s}_{1z}%
\Theta_{3,4}=-\frac{1}{2}\Theta_{3,4}\,,\\
&  \hat{s}_{2z}\Theta_{1,3}=\frac{1}{2}\Theta_{1,3}\,,\;\hat{s}_{2z}%
\Theta_{2,4}=-\frac{1}{2}\Theta_{2,4}\,,
\end{align*}
where $I$ is a $2\times2$ unity matrix, and the total spin operator is
$\mathbf{\hat{S}=\hat{s}}_{1}+\mathbf{\hat{s}}_{2}.$

The Hamiltonian of two\textit{ }interacting spins subject to the external
fields $\mathbf{G}$ and $\mathbf{F}$, respectively, is chosen as\footnote{When
restoring the Plank constant, we need to replace $J$ by $j/\hbar^{2}$ . The
factor $\hbar^{2}$ in the denominator of the interaction term ensures that $J$
has the dimension of energy.}%
\begin{align}
&  \hat{H}\left(  \mathbf{G,F,}J\right)  =2\left[  \hat{h}_{1}\otimes
I+I\otimes\hat{h}_{2}+J\mathbf{\hat{s}}_{1}\otimes\mathbf{\hat{s}}_{2}\right]
\,,\nonumber\\
&  \hat{h}_{1}=\left(  \mathbf{\mathbf{\hat{s}}\cdot G}\right)  =\frac{1}%
{2}\left(  \boldsymbol{\sigma}\mathbf{\cdot G}\right)  \,,\;\hat{h}%
_{2}=\left(  \mathbf{\mathbf{\hat{s}}\cdot F}\right)  =\frac{1}{2}\left(
\boldsymbol{\sigma}\mathbf{\cdot F}\right)  \,,\nonumber\\
&  \mathbf{\hat{s}}_{1}\otimes\mathbf{\hat{s}}_{2}=\frac{1}{4}\left[
\sigma_{1}\otimes\sigma_{1}+\sigma_{2}\otimes\sigma_{2}+\sigma_{3}%
\otimes\sigma_{3}\right]  \,, \label{eq0}%
\end{align}
where $J=J\left(  t\right)  $, in general, is a function of time, and
$\mathbf{G}=\left(  G_{1}\left(  t\right)  ,G_{2}\left(  t\right)
,G_{3}\left(  t\right)  \right)  $ and $\mathbf{F}=\left(  F_{1}\left(
t\right)  ,F_{2}\left(  t\right)  ,F_{3}\left(  t\right)  \right)  $ are, in
general, time-dependent vectors (external fields for each particle)
\cite{Fey65,Tow92}.

In the representation generated by the basis (\ref{basis}), the evolution of
the system is describes by the Schr\"{o}dinger equation%
\begin{equation}
i\frac{d\Psi}{dt}=\hat{H}\left(  \mathbf{G,F,}J\right)  \Psi\,,\;\Psi=\left(
\begin{array}
[c]{c}%
v_{1}\left(  t\right) \\
v_{2}\left(  t\right) \\
v_{3}\left(  t\right) \\
v_{4}\left(  t\right)
\end{array}
\right)  , \label{eq1}%
\end{equation}
where the Hamiltonian $\hat{H}\left(  \mathbf{G,F,}J\right)  $ is given by a
$4\times4$ matrix:%
\begin{equation}
\hat{H}=\left(
\begin{array}
[c]{cccc}%
F_{3}+G_{3}+\frac{J}{2} & F_{1}-iF_{2} & G_{1}-iG_{2} & 0\\
F_{1}+iF_{2} & G_{3}-F_{3}-\frac{J}{2} & J & G_{1}-iG_{2}\\
G_{1}+iG_{2} & J & F_{3}-G_{3}-\frac{J}{2} & F_{1}-iF_{2}\\
0 & G_{1}+iG_{2} & F_{1}+iF_{2} & \frac{J}{2}-G_{3}-F_{3}%
\end{array}
\right)  \,. \label{hamilt}%
\end{equation}
Equation (\ref{eq1}) implies for the components $v_{\mu}\left(  t\right)  $
the coupled equations%
\begin{align}
i\dot{v}_{1}  &  =\left(  F_{3}+G_{3}+\frac{J}{2}\right)  v_{1}+\left(
F_{1}-iF_{2}\right)  v_{2}+\left(  G_{1}-iG_{2}\right)  v_{3}\,,\nonumber\\
i\dot{v}_{2}  &  =\left(  F_{1}+iF_{2}\right)  v_{1}+\left(  G_{3}-F_{3}%
-\frac{J}{2}\right)  v_{2}+Jv_{3}+\left(  G_{1}-iG_{2}\right)  v_{4}%
\,,\nonumber\\
i\dot{v}_{3}  &  =\left(  G_{1}+iG_{2}\right)  v_{1}+Jv_{2}+\left(
F_{3}-G_{3}-\frac{J}{2}\right)  v_{3}+\left(  F_{1}-iF_{2}\right)
v_{4}\,,\nonumber\\
i\dot{v}_{4}  &  =\left(  G_{1}+iG_{2}\right)  v_{2}+\left(  F_{1}%
+iF_{2}\right)  v_{3}+\left(  \frac{J}{2}-G_{3}-F_{3}\right)  v_{4}\,.
\label{eq2}%
\end{align}
We call Eqs.\ (\ref{eq1}) or (\ref{eq2}) \textit{the two-spin equation}.

Let us introduce the matrices $\mathbf{\Sigma=}\mathrm{diag}\left(
\boldsymbol{\sigma}\mathbf{,}\boldsymbol{\sigma}\right)  \,$and
$\boldsymbol{\rho}=\left(  \rho_{1},\rho_{2},\rho_{3}\right)  $\textbf{,}%
\[
\rho_{1}=-\gamma^{5}=\left(
\begin{array}
[c]{lr}%
0 & I\\
I & 0
\end{array}
\right)  ,\;\rho_{2}=i\gamma^{0}\gamma^{5}\,=\left(
\begin{array}
[c]{lr}%
0 & -iI\\
iI & 0
\end{array}
\right)  ,\;\rho_{3}=\gamma^{0}=\left(
\begin{array}
[c]{lr}%
I & 0\\
0 & -I
\end{array}
\right)  \,,
\]
where the gamma-matrices are in the standard representation. Then%
\begin{equation}
\boldsymbol{\Sigma}=I\otimes\boldsymbol{\sigma}\,,\;\boldsymbol{\rho
}=\boldsymbol{\sigma}\otimes I\,,\;\left(  \boldsymbol{\Sigma}\cdot
\boldsymbol{\rho}\right)  =\boldsymbol{\sigma}\otimes\boldsymbol{\sigma}%
=\sum_{i=1}^{3}\sigma_{i}\otimes\sigma_{i}\,, \label{eq3}%
\end{equation}
so that the Hamiltonian (\ref{hamilt})\ can be written via these matrices as
follows:
\begin{equation}
\hat{H}\left(  \mathbf{G,F,}J\right)  =\left(  \boldsymbol{\rho}\mathbf{\cdot
G}\right)  +\left(  \boldsymbol{\Sigma}\mathbf{\cdot F}\right)  +\frac{J}%
{2}\left(  \boldsymbol{\Sigma}\mathbf{\cdot}\boldsymbol{\rho}\right)  .
\label{8.b}%
\end{equation}

Let us now consider a nonsingular $4\times4$\ orthogonal matrix $A$,
\begin{align}
&  A=\frac{1}{2}\left[  \mathbb{I}+\left(  \boldsymbol{\Sigma}\mathbf{\cdot
}\boldsymbol{\rho}\right)  \right]  =\left(
\begin{array}
[c]{cccc}%
1 & 0 & 0 & 0\\
0 & 0 & 1 & 0\\
0 & 1 & 0 & 0\\
0 & 0 & 0 & 1
\end{array}
\right)  ,\;\det\,A=-1\,,\nonumber\\
&  A=A^{+}=A^{-1}\,,\quad A^{2}=\mathbb{I}\,, \label{9.b}%
\end{align}
where $\mathbb{I}$ is a $4\times4$ unity matrix.\ Using the properties of the
$\sigma$-matrices,
\begin{equation}
\left[  \sigma_{i},\sigma_{j}\right]  =2i\varepsilon_{ijk}\sigma_{k}%
\,,\quad\left[  \sigma_{i},\sigma_{j}\right]  _{+}=2\delta_{ij}\,, \label{7}%
\end{equation}
where $\varepsilon_{ijk}$ is the Levi-Civita symbol ($\varepsilon_{123}=1$),
one can easily verify that%
\begin{align}
&  A\,\boldsymbol{\Sigma}\mathbf{\,}A=\boldsymbol{\rho}\mathbf{\,},\quad
A\,\boldsymbol{\rho}\mathbf{\,}A=\boldsymbol{\Sigma}\mathbf{\,},\quad
A\,\left(  \boldsymbol{\Sigma}\mathbf{\cdot}\boldsymbol{\rho}\right)
A=\left(  \boldsymbol{\Sigma}\mathbf{\cdot}\boldsymbol{\rho}\right)
\,,\label{11.b}\\
&  A\,\left(  \boldsymbol{\Sigma}+\boldsymbol{\rho}\right)
\,A=\boldsymbol{\Sigma}+\boldsymbol{\rho}\mathbf{\,},\quad A\,\left(
\boldsymbol{\Sigma}-\boldsymbol{\rho}\right)  \,A=-(\boldsymbol{\Sigma
}-\boldsymbol{\rho})\,. \label{12.b}%
\end{align}
By the use of the above properties, one can prove that
\begin{equation}
A\hat{H}\left(  \mathbf{G,F,}J\right)  \,A=\hat{H}\left(  \mathbf{F,G,}%
J\right)  \,. \label{13.b}%
\end{equation}
The latter relation implies that a solution $\Psi_{1}$ with external fields
$\mathbf{G}$ and $\mathbf{F}$ is related to a solution $\Psi_{2}$ with
external fields $\mathbf{F}$ and $\mathbf{G}$ by the matrix $A$, i.e.,
$\Psi_{2}=A\Psi_{1}$.

Let us introduce the evolution operator $R_{t}(\mathbf{G,F,}J)$ for the
two-spin equation:%
\begin{align*}
&  \Psi(t)=R_{t}(\mathbf{G,F,}J)\,\Psi_{0}\,,\;R_{0}=\mathbb{I\,},\\
&  i\frac{d}{dt}R_{t}(\mathbf{G,F,}J)=\hat{H}\left(  \mathbf{G,F,}J\right)
R_{t}(\mathbf{G,F,}J)\,.
\end{align*}
Taking (\ref{11.b}) and (\ref{12.b}) into account, one can easily see that%
\[
AR_{t}(\mathbf{G,F,}J)A=R_{t}(\mathbf{F,G,}J)\,.
\]

\subsection{Some general properties}

\begin{enumerate}
\item Let $\Psi(t)$ be a solution that corresponds to the Hamiltonian $\hat
{H}\left(  \mathbf{G,F,}J\right)  $, then $\Psi\left(  T\left(  t\right)
\right)  $, with $T\left(  t\right)  $ being a differentiable function of $t$,
is a solution that corresponds to the Hamiltonian
\[
\hat{H}\left(  \dot{T}\left(  t\right)  \mathbf{G}\left(  T\left(  t\right)
\right)  ,\dot{T}\left(  t\right)  \mathbf{F}\left(  T\left(  t\right)
\right)  ,\dot{T}\left(  t\right)  J\left(  T\left(  t\right)  \right)
\right)  ~.
\]
Thus, each solution of equation (\ref{eq1}) generates a set of solutions with
explicitly indicated arbitrariness.

\item Let the external fields be zero, $\mathbf{G}=\mathbf{F}=0$, and
$J\left(  t\right)  $ be an arbitrary function (two interacting spins without
external backgrounds). In this case, the evolution operator has the form
\begin{equation}
R_{t}\left(  0,0\mathbf{,}J\right)  =\exp\left[  i\Phi\left(  t\right)
/2\right]  \left[  \mathbb{I}\cos\Phi\left(  t\right)  -i\,A\sin\Phi\left(
t\right)  \right]  \,, \label{16.b}%
\end{equation}
where%
\begin{equation}
\Phi\left(  t\right)  =\int_{t_{0}}^{t}J\left(  \tau\right)  \,d\tau\,.
\label{14.b}%
\end{equation}

\item Let the spins do not interact, $J=0$, and the fields $\mathbf{G}\left(
t\right)  $ and$\,\mathbf{F}\left(  t\right)  $ be arbitrary. Then one can
write the expression for the evolution operator $R_{t}\left(  \mathbf{G,F,}%
0\right)  $ as follows:%
\begin{equation}
R_{t}\left(  \mathbf{G,F,}0\right)  =R_{t}\left(  \mathbf{G,}0\mathbf{,}%
0\right)  R_{t}\left(  0\mathbf{,F,}0\right)  \,. \label{16c}%
\end{equation}
In this case, the general solution of equation (\ref{eq1}) reads%
\begin{equation}
\Psi\left(  t\right)  =R_{t}\left(  \mathbf{G,}0\mathbf{,}0\right)
R_{t}\left(  0\mathbf{,F,}0\right)  \Psi_{0}=AR_{t}\left(  0\mathbf{,G,}%
0\right)  AR_{t}\left(  0\mathbf{,F,}0\right)  \Psi_{0}\,. \label{17.b}%
\end{equation}
The second form of the general solution in (\ref{17.b}) seems to be more
convenient since it is expressed via solutions of the two-spin equation with a
free first spin. This solution, in fact, is reduced to a solution of the
evolution operator for the spin equation (\ref{a.3}) because the corresponding
Hamiltonians $\left(  \boldsymbol{\Sigma}\mathbf{\cdot G}\right)  $ and
$\left(  \boldsymbol{\Sigma}\mathbf{\cdot F}\right)  $ are diagonal in this case.

\item Let the external fields be the same for both spins, $\mathbf{G}\left(
t\right)  =\mathbf{F}\left(  t\right)  $, and the interaction $J\left(
t\right)  $ be arbitrary. Taking into account the property (\ref{12.b}), one
can find the form of the evolution operator in this case:
\begin{align}
&  R_{t}\left(  \mathbf{G,G,}J\right)  =R_{t}\left(  0,0\mathbf{,}J\right)
R_{t}\left(  \mathbf{G,G,}0\right) \nonumber\\
&  \,=R_{t}\left(  0,0\mathbf{,}J\right)  \,R_{t}\left(  \mathbf{G,}%
0\mathbf{,}0\right)  R_{t}\left(  0\mathbf{,G,}0\right)  =R_{t}\left(
0,0\mathbf{,}J\right)  \,AR_{t}\left(  0\mathbf{,G,}0\right)  AR_{t}\left(
0\mathbf{,G,}0\right)  , \label{20.b}%
\end{align}
where $R_{t}\left(  0,0\mathbf{,}J\right)  $ is given by (\ref{16.b}).
Therefore, in the case under consideration, we have reduced the two-spin
problem to the one-spin problem in the\ external field $\mathbf{G}(t)$.

\item Due to the spherical symmetry, rotations commute with the interaction
operator $\left(  \boldsymbol{\Sigma}\mathbf{\cdot}\boldsymbol{\rho}\right)
=\boldsymbol{\sigma}\otimes\boldsymbol{\sigma}$ in (\ref{8.b}). Therefore, as
in the case of a single spin, the method of a \textit{rotating coordinate
system} \cite{RabRaS54}\ can be applied to the case of two interacting spins.
Let us apply, to a spinor $\psi$ that is a solution of the spin equation
(\ref{a.3}), $i\dot{\psi}=\hat{h}\psi$, the transformation $\psi^{\prime
}=r^{-1}\psi$. The spinor $\psi^{\prime}$ obeys the spin equations with a
Hamiltonian $\hat{h}^{\prime}$ given by%
\begin{equation}
\hat{h}^{\prime}=r^{-1}\hat{h}r-ir^{-1}\dot{r}\,. \label{7b}%
\end{equation}
Suppose now that $r$ is a rotation, then the operator $\mathcal{R}=r\otimes r$
commutes with $\left(  \boldsymbol{\Sigma}\mathbf{\cdot}\boldsymbol{\rho
}\right)  $. Therefore, if $\Psi$ is a solution of the two-spin equation
(\ref{eq1}), with Hamiltonian $\hat{H}$ (\ref{eq0}), the rotated spinor
$\Psi^{\prime}=\mathcal{R}^{-1}\Psi$ is a solution of this equation for a
Hamiltonian $\hat{H}^{\prime}$ given by
\begin{align}
&  \hat{H}^{\prime}=\mathcal{R}^{-1}\hat{H}\mathcal{R}-i\mathcal{R}%
^{-1}\mathcal{\dot{R}}\nonumber\\
&  =\hat{h}_{1}^{\prime}\otimes I+I\otimes\hat{h}_{2}^{\prime}+\frac{J}%
{2}\left(  \boldsymbol{\Sigma}\mathbf{\cdot}\boldsymbol{\rho}\right)  \,,
\label{7c}%
\end{align}
with the same original interaction $J$ and $\hat{h}_{i}^{\prime}$ given by
(\ref{7b}). An example of the use of this property will be given later, namely
in Eq. (\ref{33.b}), when we study a possible generalization of the Rabi
problem for two interacting spins.
\end{enumerate}

\section{Some exact solutions of two-spin equation}

\subsection{Parallel external fields}

Let the two spins be subject to different time-dependent external fields
having the same fixed direction,
\[
\mathbf{G}=\mathbf{n}B_{1}\left(  t\right)  \,,\;\mathbf{F}=\mathbf{n}%
B_{2}\left(  t\right)  \,,
\]
where $\mathbf{n}$ is a constant unity vector. Choosing an appropriated
coordinate system, or using a constant rotation, which due to (\ref{7c}) does
not change the problem, we can set $\mathbf{n}=\left(  0,0,1\right)  $, so
that
\[
\mathbf{G}=\left(  0,0,B_{1}\right)  \,,\;\mathbf{F}=\left(  0,0,B_{2}\right)
\;B_{1,2}=B_{1,2}\left(  t\right)  \mathrm{\,}.
\]
Then the Hamiltonian (\ref{8.b}) takes the form
\begin{align*}
&  i\dot{\Psi}=\hat{H}\Psi\,,\;\hat{H}=\frac{1}{2}\left[  \left(  \Sigma
_{3}+\rho_{3}\right)  B_{+}-\left(  \Sigma_{3}-\rho_{3}\right)  B_{-}%
-J\right]  +AJ,\\
&  B_{\pm}\left(  t\right)  =B_{1}\left(  t\right)  \pm B_{2}\left(  t\right)
\,.
\end{align*}
For the components $v_{1}$ and $v_{4}$ of the four-spinor $\Psi$, we get from
the Schr\"{o}dinger equation (\ref{eq1}):%
\begin{equation}
v_{1}=C_{1}\exp\left[  -i\int_{0}^{t}\left(  \frac{J}{2}+B_{+}\right)
\,d\tau\right]  \,,\;v_{4}=C_{4}\exp\left[  -i\int_{0}^{t}\left(  \frac{J}%
{2}-B_{+}\right)  \,d\tau\right]  \,,\label{3f}%
\end{equation}
with $C_{1,4}$ being complex constants. And for the components $v_{2,3}$ we
obtain the equation%
\begin{align}
&  i\dot{\psi}^{\prime}=\left[  \left(  \boldsymbol{\sigma}\mathbf{\cdot
K}\right)  -\frac{J}{2}\right]  \psi^{\prime}\,,\;\psi^{\prime}=\left(
\begin{array}
[c]{c}%
v_{2}\\
v_{3}%
\end{array}
\right)  \,,\label{3e}\\
&  \mathbf{K}\left(  t\right)  =\left(  J\left(  t\right)  ,0,B_{-}\left(
t\right)  \right)  \,.\label{3b}%
\end{align}
Doing the transformation%
\begin{equation}
\psi^{\prime}\left(  t\right)  =\exp\left[  \frac{i}{2}\int_{0}^{t}J\left(
\tau\right)  \,d\tau\right]  \psi\left(  t\right)  \,,\label{3g}%
\end{equation}
we can see that the spinor $\psi$ obeys the spin equation (\ref{a.3}) with the
external field $\mathbf{K}\left(  t\right)  $.

Exact solutions of the spin equation (\ref{a.3}) for 26 different types of
external fields of the form (\ref{3b}) are described in \cite{BagGiBL05}.
Respectively, they generate 26 sets of exact solutions of the two-spin-equation.

Below we consider two specific cases:

\begin{enumerate}
\item {\large Constant spin interaction}\newline Let $J=\varepsilon
=\mathrm{const}$. Then equations (\ref{3f}) imply%
\[
v_{1}=C_{1}e^{-i\varepsilon t/2}\exp\left(  -i\int_{0}^{t}B_{+}\,d\tau\right)
\,,\;v_{4}=C_{4}e^{-i\varepsilon t/2}\exp\left(  i\int_{0}^{t}B_{+}%
\,d\tau\right)  \,.
\]
In this case, a solution $\psi^{\prime}$\ of the problem (\ref{3e}) can be
written as%
\[
\psi^{\prime}\left(  t\right)  =\exp\left(  \frac{i\varepsilon}{2}t\right)
\psi\left(  t\right)  \,,
\]
where $\psi$ is a solution of the spin equation (\ref{a.3})\ with the field
\begin{equation}
\mathbf{K}\left(  t\right)  =\left(  \varepsilon,0,f\left(  t\right)  \right)
\,,\;f\left(  t\right)  =B_{-}\left(  t\right)  \,. \label{4}%
\end{equation}
In \cite{BagBaGS04}, one can find several functions $f\left(  t\right)  $ for
which exact solutions of the spin equation can be found.

\item {\large Fields with a constant difference}\newline Let the difference
$B_{-}=B_{1}-B_{2}=\varepsilon$ does not depend on time, $\varepsilon
=\mathrm{const}$, while the spin interaction $J$ can depend on time. It
happens sometimes in case of two interacting quantum dots \cite{EngKoLM04}.
For fields with a constant difference, a solution $\psi^{\prime}$\ of the
problem (\ref{3e}) can be written as (\ref{3g}), where $\psi$ is a solution of
the spin equation with the field
\begin{equation}
\mathbf{M}\left(  t\right)  =\left(  f\left(  t\right)  ,0,\varepsilon\right)
\,,\;f\left(  t\right)  =J\left(  t\right)  \,. \label{4a}%
\end{equation}
If $\varphi$ is a solution of the spin equation (\ref{a.3}) with the external
field $\mathbf{K}$ (\ref{4}), one can construct a solution $\psi$\ for this
equation with the external field $\mathbf{M}$ (\ref{4a}) by the transformation%
\[
\psi=\left(  2\right)  ^{-1/2}\left(  \sigma_{1}+\sigma_{3}\right)
\varphi\,.
\]
Then, we can use solutions from \cite{BagBaGS04} in order to construct exact
solutions of the two-spin equation in the fields with a constant difference.
\end{enumerate}

\section{Time-independent spin interaction and constant external fields}

Let the interaction $J$ and the fields$\,\mathbf{F}$ and$\,\,\mathbf{G}$ be
time-independent,
\begin{equation}
J\left(  t\right)  =2\gamma\,,\quad\mathbf{F}\left(  t\right)  =\mathbf{a\,}%
,\quad\mathbf{G}\left(  t\right)  =\mathbf{b\,},\label{21.b}%
\end{equation}
where $\gamma,\ \mathbf{a},\ \mathbf{b}$ are constants. In this case, we
search for solutions of two-spin equation of the form
\begin{equation}
\Psi\left(  t\right)  =\exp\left(  -i\lambda t\right)  C\,,\label{22.b}%
\end{equation}
where $C$ is a constant spinor. Substituting (\ref{22.b}) into two-spin
equation, we obtain the following equation for $C$:
\begin{equation}
D\left(  \lambda\right)  C=0\,,\;D\left(  \lambda\right)  =\gamma\left(
\boldsymbol{\Sigma}\mathbf{\cdot}\boldsymbol{\rho}\right)  +\left(
\boldsymbol{\Sigma}\mathbf{\cdot}\mathbf{a}\right)  +\left(  \boldsymbol{\rho
}\mathbf{\cdot}\mathbf{b}\right)  -\lambda\mathbb{I}\,.\label{23.b}%
\end{equation}
Its solutions exist under the following condition:
\begin{equation}
d\left(  \lambda\right)  =\det D\left(  \lambda\right)  =0.\label{24.b}%
\end{equation}
Direct calculation yields the following expression for $d\left(
\lambda\right)  $:%
\begin{align}
&  d\left(  \lambda\right)  =\lambda^{4}-2\lambda^{2}\left(  a^{2}%
+b^{2}+3\gamma^{2}\right)  +8\lambda\gamma\left[  \gamma^{2}-\left(
\mathbf{ab}\right)  \right]  -3\gamma^{4}\nonumber\\
&  \,+2\gamma^{2}\left[  a^{2}+b^{2}+4\left(  \mathbf{ab}\right)  \right]
+\left(  a^{2}-b^{2}\right)  ^{2},\quad a^{2}=\left(  \mathbf{aa}\right)
,\quad b^{2}=\left(  \mathbf{bb}\right)  ~.\label{25.b}%
\end{align}
The determinant $d\left(  \lambda\right)  $ can also be presented as%
\begin{align}
&  d\left(  \lambda\right)  =\left(  \lambda-\gamma\right)  ^{3}\left(
\lambda+3\gamma\right)  -2\left(  \lambda^{2}-\gamma^{2}\right)  \left(
a^{2}+b^{2}\right)  -8\gamma\left(  \mathbf{ab}\right)  \left(  \lambda
-\gamma\right)  +\left(  a^{2}-b^{2}\right)  ^{2}\nonumber\\
&  =\left[  \left(  \lambda+\gamma\right)  ^{2}-4\gamma^{2}-q^{2}\right]
\left[  \left(  \lambda-\gamma\right)  ^{2}-p^{2}\right]  -p^{2}q^{2}+\left(
\mathbf{pq}\right)  ^{2},\;\mathbf{p}=\mathbf{a}+\mathbf{b},\;\mathbf{q}%
=\mathbf{a}-\mathbf{b}.\label{26.b}%
\end{align}
In principle, equation (\ref{24.b}) can be solved explicitly (four roots can
be found), then equation (\ref{23.b}) allows one to find the spinor $C$.

In the particular case $\mathbf{a}=\left(  0,0,a\right)  $,$\,\mathbf{b}%
=\left(  0,0,b\right)  $, $\mathbf{p}=\left(  0,0,p=a+b\right)  $,
$\mathbf{q}=\left(  0,0,q=a-b\right)  $, the general solution of the two-spin
equation has the form%
\begin{equation}
\Psi\left(  t\right)  =\hat{U}\left(  t-t_{0}\right)  \Psi_{0}\,, \label{27.b}%
\end{equation}
where the evolution operator reads%
\begin{align}
&  2\hat{U}\left(  \tau\right)  =\left[  \left(  I+\rho_{3}\Sigma_{3}\right)
\cos{p\tau}-i\left(  \rho_{3}+\Sigma_{3}\right)  \sin{p\tau}\right]
\exp\left(  -i\gamma\tau\right) \nonumber\\
&  +\left\{  \left(  I-\rho_{3}\Sigma_{3}\right)  \cos{\omega\tau}-\left(
i/\omega\right)  \left[  q\left(  \rho_{3}-\Sigma_{3}\right)  +2\gamma\left(
\rho_{1}\Sigma_{1}+\rho_{2}\Sigma_{2}\right)  \right]  \sin{\omega\tau
}\right\}  \exp\left(  i\gamma\tau\right)  \,,\nonumber\\
&  \omega^{2}=4\gamma^{2}+q^{2}. \label{28.b}%
\end{align}

\section{Analogue of the Rabi problem for two-spin system}

\subsection{Rabi problem for one spin}

We recall that Rabi considered one spin placed in a constant magnetic field
and perpendicular to the latter a rotating field \cite{RabRaS54}. In fact, the
Rabi problem is reduced to solving spin equation (\ref{a.3}) with the external
field of the form $\mathbf{F}=\left(  A\cos\omega t,A\sin\omega t,A_{0}%
\right)  $, where $A$ and $A_{0}$ are real constants. The evolution operator
for the spin equation with the Rabi field has the form%
\begin{align}
&  \hat{u}_{F}=r_{z}\left(  \omega t\right)  \left[  I\cos\omega_{R}%
t+\frac{\left(  \alpha-ia^{\prime}\sigma_{2}\right)  ^{2}}{\alpha
^{2}+a^{\prime2}}i\sigma_{3}\sin\omega_{R}t\right]  ~,\nonumber\\
&  r_{z}\left(  \omega t\right)  =\exp\left(  -i\sigma_{3}\frac{\omega}%
{2}t\right)  ~,\ \omega_{R}^{2}=A^{2}+\left(  A_{0}-\frac{\omega}{2}\right)
^{2}~,\nonumber\\
&  a^{\prime}=\frac{A}{\omega}\,,\;a_{0}=\frac{A_{0}}{\omega}-\frac{1}%
{2}~,\ \alpha=a_{0}-\frac{\omega_{R}}{\omega}~\,,\label{r1}%
\end{align}
where $\omega_{R}$ is the \textit{Rabi frequency}. Calculating the transition
probability between two orthogonal one-spin states, we get following Rabi
\begin{equation}
\left\vert \left\langle 2\right\vert \hat{u}_{F}\left(  t\right)  \left\vert
1\right\rangle \right\vert ^{2}=\frac{\left(  \frac{\omega_{R}}{\omega
}\right)  ^{2}-a_{0}^{2}}{\left(  \frac{\omega_{R}}{\omega}\right)  ^{2}}%
\sin^{2}\omega_{R}t\,,\;\left\vert 1\right\rangle =\left(
\begin{array}
[c]{c}%
1\\
0
\end{array}
\right)  ,\;\left\vert 2\right\rangle =\left(
\begin{array}
[c]{c}%
0\\
1
\end{array}
\right)  .\label{r2}%
\end{equation}
Since $\left(  \omega_{R}/\omega\right)  ^{2}\geq a_{0}^{2}$, the amplitude of
the probability has a maximum in the resonance frequency $\omega=2A_{0}$ of
external field.

If we consider two noninteracting spins, one of them free and another one
placed in the Rabi field, we can describe such a system by the two-spin
equation with $\mathbf{G}=J=0$ and $\mathbf{F}=\left(  A\cos\omega
t,A\sin\omega t,A_{0}\right)  $, and the Hamiltonian
\begin{equation}
\hat{H}=\left(  \boldsymbol{\Sigma}\mathbf{\cdot}\mathbf{F}\right)
\,.\label{r0}%
\end{equation}
The evolution operator for the two-spin equation (\ref{r0}) in such a case can
be written, in virtue of relations (\ref{eq3}), as follows:%
\begin{align}
&  R_{t}\left(  0\mathbf{,F,}0\right)  =I\otimes\hat{u}_{F}=\exp\left(
-i\Sigma_{3}\frac{\omega}{2}t\right)  R_{\Sigma}\left(  \omega_{R}t\right)
\,,\nonumber\\
&  R_{\Sigma}\left(  \omega_{R}t\right)  =\left[  \mathbb{I}\cos\omega
_{R}t+\frac{\left(  \alpha-ia^{\prime}\Sigma_{2}\right)  ^{2}}{\alpha
^{2}+a^{\prime2}}i\Sigma_{3}\sin\omega_{R}t\right]  \,.\label{r2b}%
\end{align}

\subsection{Possible generalizations of the Rabi problem for two interacting
spins}

\subsubsection{Different Rabi fields for each spin}

Let us considered two interacting spins each of them placed in a Rabi field
and with a spin interaction that does not depend on time. Such a situation is
described by the two-spin equation (\ref{eq1}) with
\begin{align}
&  J\left(  t\right)  =J,\ \mathbf{F}=\left(  A\cos\left(  \omega
t+\varphi_{1}\right)  ,A\sin\left(  \omega t+\varphi_{1}\right)
,A_{0}\right)  ~,\nonumber\\
&  \mathbf{G}=\left(  B\cos\left(  \omega t+\varphi_{2}\right)  ,B\sin\left(
\omega t+\varphi_{2}\right)  ,B_{0}\right)  ~,\label{31.b}%
\end{align}
where $J,\,A,\,B,\,A_{0},\,B_{0},\,\varphi_{1},\,\varphi_{2},\,\omega$ are
constants. Its solution can be chosen in the form%

\begin{align}
&  \Psi\left(  t\right)  =\exp\left(  -i\lambda\omega t\right)  \mathcal{R}%
_{z}\left(  \omega t\right)  C\,,\nonumber\\
&  \mathcal{R}_{z}\left(  \omega t\right)  =\exp\left[  -i\frac{\omega t}%
{2}\left(  \Sigma_{3}+\rho_{3}\right)  \right]  \,. \label{33.b}%
\end{align}
where $C$, with components $C_{k}\,(k=1,2,3,4)$, is a constant bispinor. Note
that $\mathcal{R}_{z}$ is a rotation in the $z$-direction by the angle $\omega
t$, such that we use a \textit{rotating coordinate system} that rotates with
the field, similar to ordinary Rabi problem. Taking the Hamiltonian $\hat{H}%
$\ with the fields (\ref{31.b}), the rotation $\mathcal{R}_{z}$ obeying
(\ref{7c}), and setting $D\left(  \lambda\right)  =\hat{H}^{\prime}%
-\lambda\mathbb{I}$, we see that the bispinor $C$ obeys a linear set of
equations%
\begin{equation}
D(\lambda)\,C=0, \label{34.b}%
\end{equation}
where the \textit{constant} matrix $D(\lambda)$ has the form (\ref{23.b})
with
\begin{align}
&  \gamma=\frac{J}{2\omega}\,,\;\mathbf{a}=\left(  a^{\prime}\cos\varphi
_{1},a^{\prime}\sin\varphi_{1},a_{0}\right)  \,,\;a^{\prime}=\frac{A}{\omega
}\,,\;a_{0}=\frac{A_{0}}{\omega}-\frac{1}{2}~,\ \nonumber\\
&  \mathbf{b}=\left(  b^{\prime}\cos\varphi_{2},b^{\prime}\sin\varphi
_{2},b_{0}\right)  \,,\;\,b^{\prime}=\frac{B}{\omega},\,b_{0}=\frac{B_{0}%
}{\omega}-\frac{1}{2}\,. \label{35.b}%
\end{align}
Thus, the problem under consideration is reduced to a problem for two spins in
time-independent fields and with a constant interaction.

\subsubsection{Equal Rabi fields for both spins}

Let us considered two interacting spins each of them placed in the same Rabi
field and with a time-dependent spin interaction $J\left(  t\right)  $. Such a
situation is described by two-spin equation (\ref{eq1}) with%
\[
\mathbf{G}=\mathbf{F}=\left(  A\cos\omega t,A\sin\omega t,A_{0}\right)
\,,\;\gamma\left(  t\right)  =\frac{1}{2\omega}\int_{t_{0}}^{t}J\left(
\tau\right)  \,d\tau.
\]
In the case under consideration, the evolution operator has the form
(\ref{20.b})
\begin{align*}
R_{t}\left(  \mathbf{F,F,}J\right)   &  =R_{t}\left(  0,0\mathbf{,}%
\gamma\right)  R_{t}\left(  \mathbf{F,}0\mathbf{,}0\right)  R_{t}\left(
0\mathbf{,F,}0\right)  \,,\\
R_{t}\left(  0,0\mathbf{,}\gamma\right)   &  =\exp\left[  -i\left(
\boldsymbol{\Sigma}\mathbf{\cdot}\boldsymbol{\rho}\right)  \omega\gamma\left(
t\right)  \right]  \,,
\end{align*}
with $R_{t}\left(  0\mathbf{,F,}0\right)  $ given by (\ref{r2b}) and%
\begin{align*}
&  R_{t}\left(  \mathbf{F,}0\mathbf{,}0\right)  =\exp\left(  -i\rho_{3}%
\frac{\omega}{2}t\right)  R_{\rho}\left(  \omega_{R}t\right)  ~,\\
&  R_{\rho}\left(  \omega_{R}t\right)  =\left[  \mathbb{I}\cos\omega
_{R}t+\frac{\left(  \alpha-ia^{\prime}\rho_{2}\right)  ^{2}}{\alpha
^{2}+a^{\prime2}}i\rho_{3}\sin\omega_{R}t\right]  \,.
\end{align*}
Using the commutation relations between rotations and the operator $\left(
\boldsymbol{\Sigma}\mathbf{\cdot}\boldsymbol{\rho}\right)  $, and the fact
that $\left[  \Sigma_{i},\rho_{j}\right]  =0,$ we find%
\begin{align*}
&  R_{t}\left(  \mathbf{F,F,}J\right)  =\mathcal{R}_{z}\left(  \omega
t\right)  \exp\left[  -i\left(  \mathbf{\Sigma\cdot}\mathbf{\rho}\right)
\omega\gamma\left(  t\right)  \right]  R_{\rho}\left(  t\right)  R_{\Sigma
}\left(  t\right)  \,,\\
&  \mathcal{R}_{z}\left(  \omega t\right)  =r_{z}\otimes r_{z}=\exp\left[
-i\left(  \rho_{3}+\Sigma_{3}\right)  \frac{\omega}{2}t\right]  \,.
\end{align*}

In the absence of the circular field $\left(  A=0\right)  $, the corresponding
stationary states $\left\vert \Psi_{i}\right\rangle $ and energy eigenvalues
$\lambda_{i}$\ of the problem are:
\begin{align}
&  \left\vert \Psi_{1}\right\rangle =\left\vert 1\right\rangle \otimes
\left\vert 1\right\rangle \,,\;\lambda_{1}=\frac{J}{2}+2A_{0}\,,\nonumber\\
&  \left\vert \Psi_{2}\right\rangle =\frac{1}{\sqrt{2}}\left(  \left\vert
1\right\rangle \otimes\left\vert 2\right\rangle +\left\vert 2\right\rangle
\otimes\left\vert 1\right\rangle \right)  \,,\;\lambda_{2}=\frac{J}%
{2}\,,\nonumber\\
&  \left\vert \Psi_{3}\right\rangle =\frac{1}{\sqrt{2}}\left(  \left\vert
2\right\rangle \otimes\left\vert 1\right\rangle -\left\vert 1\right\rangle
\otimes\left\vert 2\right\rangle \right)  \,,\;\lambda_{3}=-\frac{3}%
{2}J\,,\nonumber\\
&  \left\vert \Psi_{4}\right\rangle =\left\vert 2\right\rangle \otimes
\left\vert 2\right\rangle \,,\;\lambda_{4}=\frac{J}{2}-2A_{0}\,. \label{r3b}%
\end{align}
Using the fact that $\left\vert \Psi_{i}\right\rangle $ are eigenvectors of
$\left(  \boldsymbol{\Sigma}\mathbf{\cdot}\boldsymbol{\rho}\right)  $ with the
eigenvalues $\lambda_{1,2,4}=1/2$ and $\lambda_{3}=-3/2$, we obtain:

\begin{description}
\item[a) ] $\left\langle \Psi_{3}\right\vert R_{t}\left(  \mathbf{F,F,}%
J\right)  \left\vert \Psi_{k}\right\rangle =0\,,k=1,2,4~,$ which is, in fact,
a consequence of the conservation of the total angular momentum.

\item[b) ]
\begin{align*}
&  \left\langle \Psi_{4}\right\vert R_{t}\left(  \mathbf{F,F,}J\right)
\left\vert \Psi_{1}\right\rangle =\exp\left(  -i\omega\gamma\right)
\left\langle 2\right\vert \hat{u}_{F}\left\vert 1\right\rangle ^{2}\ ,\\
&  \left\vert \left\langle \Psi_{4}\right\vert R_{t}\left(  \mathbf{F,F,}%
J\right)  \left\vert \Psi_{1}\right\rangle \right\vert ^{2}=\left[
\frac{\left(  \frac{\omega_{R}}{\omega}\right)  ^{2}-a_{0}^{2}}{\left(
\frac{\omega_{R}}{\omega}\right)  ^{2}}\sin^{2}\omega_{R}t\right]  ^{2}\,,
\end{align*}
where $\left\vert \left\langle 2\right\vert \hat{u}_{F}\left\vert
1\right\rangle \right\vert ^{2}$\ is given by (\ref{r2}).

\item[c) ]
\begin{align}
&  \left\langle \Psi_{2}\right\vert R_{t}\left(  \mathbf{F,F,}J\right)
\left\vert \Psi_{1}\right\rangle =\frac{2}{\sqrt{2}}\exp\left(  -i\omega
\gamma\right)  \left\langle 1\right\vert \hat{u}_{F}\left\vert 1\right\rangle
\left\langle 2\right\vert \hat{u}_{F}\left\vert 1\right\rangle \,,\nonumber\\
&  \left\langle \Psi_{2}\right\vert R_{t}\left(  \mathbf{F,F,}J\right)
\left\vert \Psi_{4}\right\rangle =\frac{2}{\sqrt{2}}\exp\left(  -i\omega
\gamma\right)  \left\langle 1\right\vert \hat{u}_{F}\left\vert 2\right\rangle
\left\langle 2\right\vert \hat{u}_{F}\left\vert 2\right\rangle \,. \label{r3}%
\end{align}
The one-spin amplitudes involved in (\ref{r3}) can be obtained from
(\ref{r1}),%
\begin{align*}
&  \left\langle 2\right\vert \hat{u}_{F}\left(  t\right)  \left\vert
1\right\rangle =e^{i\frac{\omega}{2}t}\frac{2a^{\prime}\alpha}{\alpha
^{2}+a^{\prime2}}\sin\omega_{R}t\ ,\\
&  \left\langle 1\right\vert \hat{u}_{F}\left(  t\right)  \left\vert
1\right\rangle =e^{i\frac{\omega}{2}t}\left(  \cos\omega_{R}t-i\omega
\frac{a_{0}}{\omega_{R}}\sin\omega_{R}t\right)  \,,\\
&  \left\langle 2\right\vert \hat{u}_{F}\left(  t\right)  \left\vert
2\right\rangle =e^{-i\frac{\omega}{2}t}\left(  \cos\omega_{R}t+i\omega
\frac{a_{0}}{\omega_{R}}\sin\omega_{R}t\right)  \,.
\end{align*}

\end{description}

We see that the transition from the state $\left\vert \Psi_{1}\right\rangle $
to the one $\left\vert \Psi_{4}\right\rangle $ has a resonance at
$\omega=\omega_{1}=2A_{0}$,%
\[
\max\left(  \left\vert \left\langle \Psi_{1}\right\vert R_{t}\left\vert
\Psi_{4}\right\rangle \right\vert ^{2}\right)  =1\,.
\]
Yet, for an external field in the frequency $\omega=\omega_{1}$, transitions
from the state $\left\vert \Psi_{2}\right\rangle $ to $\left\vert \Psi
_{1}\right\rangle $ and $\left\vert \Psi_{4}\right\rangle $ have resonances,%
\[
\max\left(  \left\vert \left\langle \Psi_{2}\right\vert R_{t}\left\vert
\Psi_{1}\right\rangle \right\vert ^{2}\right)  =\max\left(  \left\vert
\left\langle \Psi_{2}\right\vert R_{t}\left\vert \Psi_{4}\right\rangle
\right\vert ^{2}\right)  =1/2\,,
\]

In addition, these transitions have resonances at $\omega=\omega_{2}=2\left(
A_{0}-A\right)  $,%
\[
\max\left(  \left|  \left\langle \Psi_{2}\right|  R_{t}\left|  \Psi
_{1}\right\rangle \right|  ^{2}\right)  =\max\left(  \left|  \left\langle
\Psi_{2}\right|  R_{t}\left|  \Psi_{4}\right\rangle \right|  ^{2}\right)
=1/2\,.
\]

\begin{acknowledgement}
V.G.B. thanks grant SS-5103.2006.2 of the President of Russia and RFBR grant
06-02-16719 for partial support; M.C.B. thanks FAPESP; D.M.G. thanks FAPESP
and CNPq for permanent support.
\end{acknowledgement}


\begin{thebibliography}{99}                                                                                               %


\bibitem {Nus73}Nussenzveig H.M., \emph{Introduction to Quantum Optics},
(Gordon and Breach, New York 1973)

\bibitem {AllEb75}Allen L., Eberly J.H., \emph{Optical Resonance and Two-Level
Atoms}, (Wiley, NY 1975)

\bibitem {RabRaS54}Rabi I.I., Ramsey N.F., Schwinger J., \emph{Use of Rotating
Coordinates in Magnetic Resonance Problems}, Rev. Mod. Phys. \textbf{26}, 167 (1945)

\bibitem {FeyVe57}Feynman R.P., Vernon F.L., \emph{Geometrical Representation
of the Schr\"{o}dinger Equation for Solving Maser Problems}, J. Applied
Physics \textbf{28}, 49 (1957)

\bibitem {Fey65}Feynman R.P., \emph{Lectures on Physics}, Vol. III, Lecture
12, Addison-Wesley Publ. Co. (1965)

\bibitem {Tow92}Townsend J.S., \emph{A Modern Approach to Quantum Mechanics}
(McGraw-Hill, New York 1992)

\bibitem {QuantComp}Feynman R.P., \emph{Simulating Physics with Computers},
Int. J. Theor. Phys. \textbf{21} 467 (1982); Deutsch D., \emph{Quantum Theory,
the Church-Turing Principle and the Universal Quantum Computer}, Proc. R. Soc.
London \textbf{A400}, 97 (1985); Feynman R.P., \emph{Quantum Mechanical
Computers}, Found. Phys. \textbf{16}, 507 (1986); Valiev K.A., Kokin A.A.,
\emph{Quantum computers: Hopes and reality}, (NITS, Ijevsk: 2001); Berman G.,
Doolen G., Mainieri R., Tsifrinovich V., \emph{Introduction to Quantum
Computers}, (World Scientific, Singapore 1999); Nielsen M.A., Chuang I.L.,
\emph{Quantum Computation and Quantum Information}, (Cambridge University
Press, Cambridge 2000)

\bibitem {BagGiBL05}Bagrov V.G., Gitman D.M., Baldiotti M.C., and Levin A.,
\emph{Spin equation and its solutions, }Ann. Phys. (Leipzig) \textbf{14}
[11-12] (2005) pp.764-789

\bibitem {Pau27}Pauli W., \emph{Zur Quantenmechanik des magnetischen
Elektrons}, Zeit. Phys. \textbf{43}, 601 (1927)

\bibitem {BloSi40}Bloch F., Siegert A., \emph{Magnetic Resonance for
Nonrotating Fields}, Phys. Rev. \textbf{57}, 522 (1940)

\bibitem {ZakSh84}Novikov S. et al, \emph{Theory of Solitons}, (Consultants
Bureau, New York, London 1984)

\bibitem {Rab37}Rabi I.I., \emph{Space Quantization in a Gyrating Magnetic
Field}, Phys. Rev. \textbf{51}, 652 (1937)

\bibitem {BagBaGW01}Bagrov V.G., Barata J.C., Gitman D.M., Wreszinski W.F.,
\emph{Aspects of two-level systems under external time-dependent fields}, J.
Phys A\textbf{34}, 10869 (2001)

\bibitem {BagBaGS04}Bagrov V.G., Baldiotti M.C., Gitman D.M., Shamshutdinova
V.V., \emph{Darboux transformation for two-level system,} Ann. Phys. (Leipzig)
\textbf{14}, No 6, 390 (2005)

\bibitem {BarCo02}Barata J.C., Cortez D.A., \emph{Time evolution of two-level
systems driven by periodic fields}, Phys Lett. A \textbf{301}, 350 (2002)

\bibitem {Bar00}Barata. J. C., \emph{On Formal Quasi-Periodic Solutions of the
Schr\"{o}dinger Equation for a Two-Level System with a Hamiltonian Depending
Quasi-Periodically in Time}, Rev. Math. Phys. \textbf{12}, 25 (2000)

\bibitem {EngKoLM04}Engel H., Kouwenhoven L.P., Loss D., Marcus C.M.,
\emph{Controlling Spin in Quantum Dots}, arXive e-print cond-mat/0409294v1 (2004)
\end{thebibliography}
\end{document}